\RequirePackage[mathlines]{lineno} 
\documentclass[a4paper,11pt]{article}
\pdfoutput=1 

\usepackage{jheppub} 

\usepackage[T1]{fontenc} 

\usepackage{threeparttable}
\usepackage{multirow}
\usepackage{subfigure}
\usepackage{hyperref}
\usepackage{float}
\let\oldequation\equation
\let\oldendequation\endequation
\renewenvironment{equation}
  {\linenomathNonumbers\oldequation}
  {\oldendequation\endlinenomath}

\def \jpsi {J/\psi}

\def \gev  {~\mbox{GeV}}
\def \gevc {~\mbox{GeV/$c$}}
\def \gevcc{~\mbox{GeV/$c^2$}}
\def \mev  {~\mbox{MeV}}

\begin{document}

\title{\boldmath Search for the charmonium weak decays $\jpsi\to D_{s}^{-}\rho^{+}+c.c.$ and $\jpsi\to D_{s}^{-}\pi^{+}+c.c.$}
\collaborationImg{\includegraphics[height=30mm,angle=90]{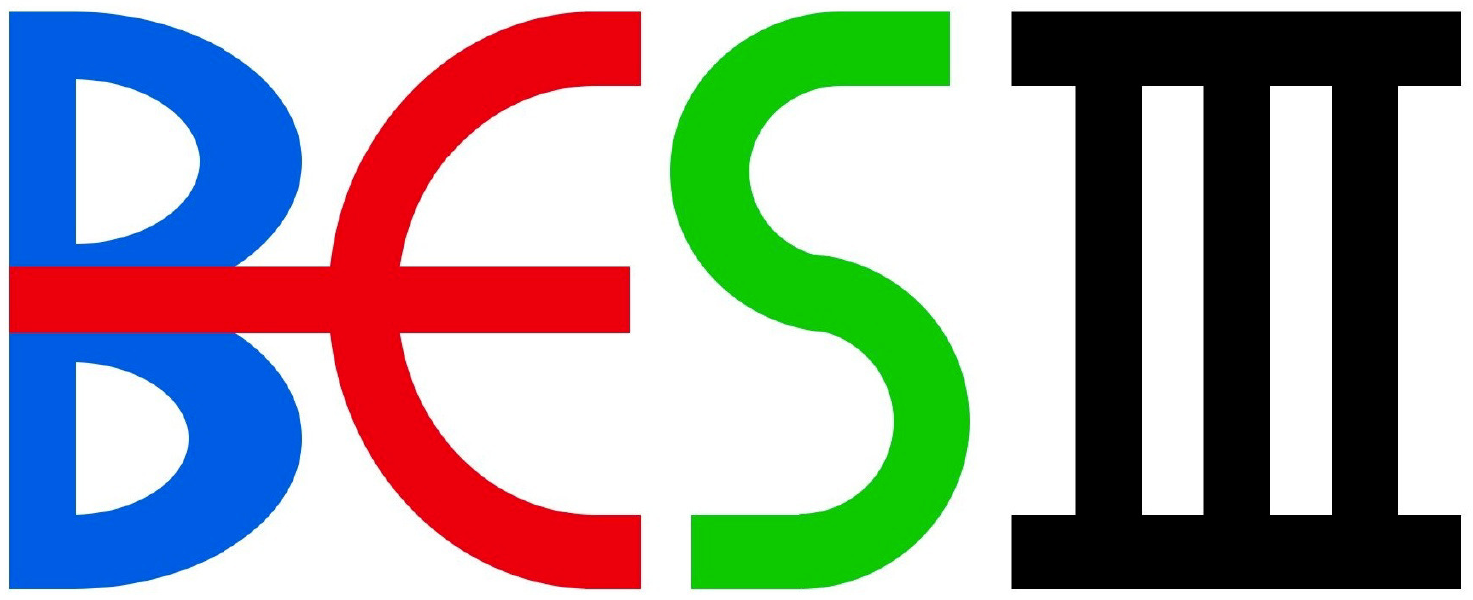}}
\collaboration{The BESIII collaboration}
\emailAdd{besiii-publications@ihep.ac.cn}

\abstract{
Based on $(10087\pm44)\times 10^6$ $\jpsi$ events recorded with the BESIII detector, we search for the rare charmonium weak decays $\jpsi\to D_{s}^{-}\rho^{+}+c.c.$ and $\jpsi\to D_{s}^{-}\pi^{+}+c.c.$  No signal is observed, and upper limits on the branching fractions at the $90\%$ confidence level are set as $\mathcal{B}(\jpsi\to D_{s}^{-}\rho^{+}+c.c.)<8.0\times10^{-7}$ and $\mathcal{B}(\jpsi\to D_{s}^{-}\pi^{+}+c.c.)<4.1\times10^{-7}$. Our results provide the most stringent experimental constraints on these decays.
}

\keywords{BESIII, charmonium, weak decay}

\arxivnumber{}

\newcommand{\BESIIIorcid}[1]{\href{https://orcid.org/#1}{\hspace*{0.1em}\raisebox{-0.45ex}{\includegraphics[width=1em]{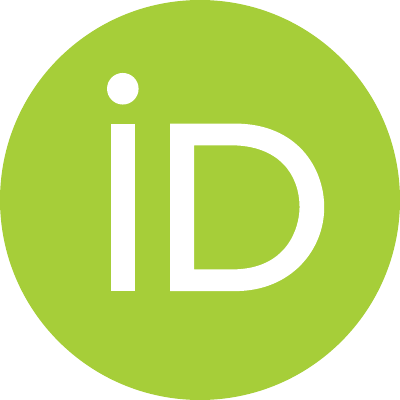}}}}
\maketitle
\flushbottom


\section{INTRODUCTION}
\label{sec:introduction}
\hspace{1.5em} 
As the mass of the $\jpsi$ lies below the $D\Bar{D}$ threshold, its decay to $D\Bar{D}$ is forbidden. However, weak decays of the $\jpsi$ to a charmed meson $D$ or $D_{s}$ accompanied by a light meson, such as a $\rho$ or $\pi$, are allowed in the Standard Model~(SM). Compared with the extensively studied decays of $\jpsi$ caused by the strong and electromagnetic interactions, weak decays of the $\jpsi$ are rarely mentioned. The inclusive branching fraction~(BF) of charmonium rare weak decays is predicted to be of the order of $10^{-8}$ in the SM~\cite{verma:1990, Sanchis:1994, Sanchis:1993, sharma:1999, wang:2008a, wang:2008b, shen:2008, dhir:2013, ivanov:2015, tian:2017, Sun:2023uyn, Meng:2024nyo}. The BFs of exclusive decays $J/\psi\to D_{s}^{-}\rho^{+}$ and $J/\psi\to D_{s}^{-}\pi^{+}$ are predicted to be $(12.6-51.1)\times10^{-10}$ and $(2.0-7.5)\times10^{-10}$~\cite{wang:2008a, shen:2008, dhir:2013, Sun:2023uyn} in the SM, respectively, as shown in Table~\ref{tab:prediction}. Figure~\ref{fig:feynman} shows an example of the Feynman diagram at the tree level for the charmonium SM weak decays $\jpsi\to D_{s}^{-}\rho^{+}$ and $\jpsi\to D_{s}^{-}\pi^{+}$. These $\jpsi$ weak decays are mediated via the spectator mechanism in which one of the charm~(anti-charm) quarks decays and the other does not ({\it i.e.}, is a spectator). The ratio between the expectation of the $D_s$ channel and the $D$ channel $\frac{\mathcal{B}_{SM}(J/\psi\to D_{s}^{-}\rho^{+}(\pi^{+}))}{\mathcal{B}_{SM}(J/\psi\to D^{-}\rho^{+}(\pi^{+}))}$ is approximately 20 because the $D_s$ channel is Cabibbo favored within the SM~\cite{wang:2008a, dhir:2013, Sun:2023uyn}. The decays $\jpsi\to D_{s}^{-}\rho^{+}$ and $\jpsi\to D_{s}^{-}\pi^{+}$ are two of the charmonium weak decays involving a $D_{(s)}$ meson that have the largest BFs expected in the SM\cite{Sun:2023uyn}, making them the most sensitive for observing a SM signal.
\begin{table*}[!htbp]
\caption{Theoretical predictions for the BFs of the weak decays $\jpsi\to D_{s}^{-}{\rho}^{+}$ and $\jpsi\to D_{s}^{-}{\pi}^{+}$ within the SM, where QCDSR is the QCD sum rule model, BSW is the Bauer-Stech-Wirbel model, and CLFQM is the covariant light-front quark model.}
\setlength{\abovecaptionskip}{1.2cm}
\setlength{\belowcaptionskip}{0.2cm}
\begin{center}
\footnotesize
\vspace{-0.0cm}
\begin{tabular}{l|cccccccccc}
\hline \hline
                    Model & QCDSR~\cite{wang:2008a}
                    & BSW~\cite{dhir:2013}
                    & CLFQM~\cite{shen:2008} & CLFQM~\cite{Sun:2023uyn}\\ 
                    \hline
                    $\mathcal{B}(\jpsi\to D_{s}^{-}\rho^{+})$ $(\times10^{-10})$ 
                    & $12.6_{-1.2}^{+3.0}$
                    & $51.1_{-6.0}^{+7.6}$
                    & $28_{-9}^{+0}$ 
                    & $29.5_{-0.5-1.4-1.9}^{+0.6+1.1+1.5}$ 
                    \\
                    $\mathcal{B}(\jpsi\to D_{s}^{-}\pi^{+})$ $(\times10^{-10})$ 
                    & $2.0_{-0.2}^{+0.4}$
                    & $7.41_{-0.23}^{+0.13}$
                    & $2.5_{-0.1}^{+0.0}$
                    & $3.64_{-0.06-0.38-0.96}^{+0.06+0.34+0.78}$
                    \\
\hline \hline
\end{tabular}
\label{tab:prediction}
\end{center}
\end{table*}
\vspace{-0.0cm}
\vspace{-0.0cm}
\begin{figure}[htbp] \centering
	\setlength{\abovecaptionskip}{-1pt}
	\setlength{\belowcaptionskip}{10pt}
	\includegraphics[width=7.0cm]{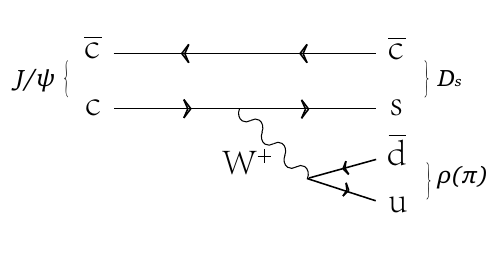}
	\caption{Feynman diagram for both $\jpsi\to D_{s}^{-}\rho^{+}$ and $\jpsi\to D_{s}^{-}\pi^{+}$ decays at the tree-level.}
	\label{fig:feynman}
\end{figure}
\vspace{-0.0cm}

Many hadronic weak decays of the $\jpsi$ have been studied in theory, such as several branching ratios for $\jpsi\to SS, VS$ decays~($S$ and $V$ represent a scalar, or a pseudoscalar, and a vector meson, respectively). The decay rates of $\jpsi\to D_{s} M$~($M$ represents a meson) are given by
\begin{eqnarray}
\Gamma_{\jpsi\to D_{s}\it{M}}= \frac{1}{3} \frac{1}{8\pi} | \mathcal{A}(\jpsi\to D_{s}M )|^{2} \frac{|\vec{p}_{D_{s}}|}{m_{\jpsi}^{2}}, 
\end{eqnarray}
where $\vec{p}_{D_{s}}$ is the three-momentum of the $D_{s}$ meson, the factor $\frac{1}{3}$ is from the spin average of the $\jpsi$, and $\mathcal{A}(\jpsi\to D_{s}M)$ is the decay amplitude which can be calculated based on the three-point QCD sum rules~(QCDSR)~\cite{wang:2008a}. The BF ratio for the hadronic weak decays $\jpsi\to D_{s}^{-}\rho^{+}$ and $\jpsi\to D_{s}^{-}\pi^{+}$ is predicted to be approximately 4.2 by using a factorization scheme~\cite{sharma:1999}. The experimental results of known charmonium rare weak decays are shown in Table~\ref{tab:exp}~\cite{Ablikim:2019hff, Li:2024moj, bes:2008, bes3:2014xbo, bes3:2014, bes3:2017, bes3:2021, BESIII:2023fqz, BESIII:2022ibp, BESIII:2023qpx, BESIII:2025wsu}.

Meanwhile, various new physics models beyond the SM, including the top-color model~\cite{hill:1995}, the minimal super-symmetric SM with or without R-parity~\cite{Aulakh:1982yn}, and the two-Higgs doublet model~\cite{Glashow:1976nt} allow the $\jpsi$ weak decay inclusive BFs to be enhanced up to $10^{-5}$~\cite{Datta:1998yq}. Most new physics models enhance $\jpsi$ weak decays either via flavor-changing neutral-current~(FCNC) processes or through the exchange of a charged boson. Furthermore, several models, such as the two-Higgs doublet model,  predict a larger enhancement for the $D_s$ channel than for the $D$ channel~\cite{Pich:2009sp}. The processes $\jpsi\to D_{s}^{-}\rho^{+}$ and $\jpsi\to D_{s}^{-}\pi^{+}$ provide a unique opportunity to search for new physics beyond the SM~\cite{Ablikim:2019hff, Li:2024moj}.  If a signal for either of these two decays is observed with BFs in the range of $10^{-8}$ to $10^{-6}$, it would indicate new physics beyond the SM.
\begin{table*}[!htbp]
\caption{Summary of studies of charmonium weak decays, listing the decay mode, the total number of $\jpsi$ or $\psi(3686)$ events, and the upper limits~(ULs) on the BFs at the 90\% confidence level~(C.L.).}
\setlength{\abovecaptionskip}{1.2cm}
\setlength{\belowcaptionskip}{0.2cm}
\begin{center} 
\footnotesize
\vspace{-0.0cm}
\begin{tabular}{l|llll}
\hline \hline
		Experiment & Decay (+c.c) & $N_{\jpsi}$ or $N_{\psi(3686)}$ & UL at 90\% C.L. & Year\\
		\hline
		 BESII & $J/\psi\to D_s^-\pi^{+}$ & $58\times10^{6}$ & $1.4\times10^{-4}$ & 2008 \cite{bes:2008}\\
		 BESII & $J/\psi\to D^0K^0$ & $58\times10^{6}$ & $1.7\times10^{-4}$ & 2008 \cite{bes:2008}\\
		 BESIII & $J/\psi\to D_s^-\rho^+$ & $225\times10^{6}$ & $1.3\times10^{-5}$ & 2014 \cite{bes3:2014xbo}\\
		 BESIII & $J/\psi\to D^0K^{*0}$ & $225\times10^{6}$ & $2.5\times10^{-6}$ & 2014 \cite{bes3:2014xbo}\\
		BESIII & $J/\psi\to D_s^-e^+\nu_e$ & $225\times10^{6}$ & $1.3\times10^{-6}$ & 2014 \cite{bes3:2014}\\
		BESIII & $J/\psi\to D_s^{*-}e^+\nu_e$ & $225\times10^{6}$ & $1.8\times10^{-6}$ & 2014 \cite{bes3:2014}\\
		BESIII & $J/\psi\to D^0e^+e^-$ & $1310.6\times10^{6}$ & $8.5\times10^{-8}$ & 2017 \cite{bes3:2017}\\
        BESIII & $\psi(3686)\to D^0e^+e^-$ & $1310.6\times10^{6}$ & $1.4\times10^{-7}$ & 2017 \cite{bes3:2017}\\
		BESIII & $J/\psi\to D^-e^+\nu_e$ & $10087\times10^{6}$ & $7.1\times10^{-8}$ & 2021 \cite{bes3:2021}\\
        BESIII & $\psi(3686)\to \Lambda_{c}^{+}\bar{\Sigma}^{-}$ & $448.1\times10^{6}$ & $4.7\times10^{-7}$ & 2023 \cite{BESIII:2022ibp}\\
            BESIII & $J/\psi\to D^-\mu^+\nu_\mu$ & $10087\times10^{6}$ & $5.6\times10^{-7}$ & 2024 \cite{BESIII:2023fqz}\\
            BESIII & $J/\psi\to D^-\rho^+$ & $10087\times10^{6}$ & $6.0\times10^{-7}$ & 2024 \cite{BESIII:2023qpx}\\
         BESIII & $J/\psi\to D^-\pi^+$ & $10087\times10^{6}$ & $7.0\times10^{-8}$ & 2024 \cite{BESIII:2023qpx}\\
         BESIII & $J/\psi\to \bar{D}^0\rho^0$ & $10087\times10^{6}$ & $5.2\times10^{-7}$ & 2024 \cite{BESIII:2023qpx}\\
         BESIII & $J/\psi\to \bar{D}^0\eta$ & $10087\times10^{6}$ & $6.8\times10^{-7}$ & 2024 \cite{BESIII:2023qpx}\\
         BESIII & $J/\psi\to \bar{D}^0\pi^0$ & $10087\times10^{6}$ & $4.7\times10^{-7}$ & 2024 \cite{BESIII:2023qpx}\\
         BESIII & $J/\psi\to D^0\mu^+\mu^-$ & $10087\times10^{6}$ & $1.1\times10^{-7}$ & 2025 \cite{BESIII:2025wsu}\\
            
\hline \hline
\end{tabular}
\label{tab:exp}
\end{center}
\end{table*}
\vspace{-0.0cm}

In this paper, we search for the charmonium weak decays  $\jpsi\to D_{s}^{-}\rho^{+}$ and $\jpsi\to D_{s}^{-}\pi^{+}$ using $(10087\pm44)\times 10^6$ $\jpsi$ events collected at the BESIII detector~\cite{bes3:totJpsiNumber}. Throughout this paper, charge-conjugate processes are always implied unless explicitly specified.

\section{BESIII DETECTOR AND MONTE CARLO SIMULATION}
\label{sec:detector}
\hspace{1.5em} 
The BESIII detector~\cite{bes3:2010detector} records symmetric $e^+e^-$ collisions provided by the BEPCII storage ring~\cite{Yu:IPAC2016-TUYA01} in the center-of-mass energy range from 1.84 to 4.95~GeV, with a peak luminosity of $1.1 \times 10^{33}\;\text{cm}^{-2}\text{s}^{-1}$ achieved at $\sqrt{s} = 3.773\;\text{GeV}$. BESIII has collected large data samples in this energy region~\cite{Ablikim:2019hff, EcmsMea, EventFilter, Liao:2025lth}. The cylindrical core of the BESIII detector covers 93\% of the full solid angle and consists of a helium-based multilayer drift chamber~(MDC), a plastic scintillator time-of-flight system~(TOF), and a CsI(Tl) electromagnetic calorimeter~(EMC), which are all enclosed in a superconducting solenoidal magnet providing a 1.0~T magnetic field. The magnetic field was 0.9~T in 2012, which affects 11\% of the total $J/\psi$ data. The solenoid is supported by an octagonal flux-return yoke with resistive plate counter muon identification modules interleaved with steel. 
The charged-particle momentum resolution at $1~{\rm GeV}/c$ is $0.5\%$, and the ${\rm d}E/{\rm d}x$ resolution is $6\%$ for electrons from Bhabha scattering. The EMC measures photon energies with a resolution of $2.5\%$ ($5\%$) at $1$~GeV in the barrel (end-cap) region. The time resolution in the TOF barrel region is 68~ps, while that in the end-cap region is 110~ps. The end-cap TOF system was upgraded in 2015 using multigap resistive plate chamber technology, providing a time resolution of 60~ps, for 87\% of the data used in this analysis~\cite{etof, etof2, etof3}.

Monte Carlo (MC) simulated data samples produced with a {\sc geant4}-based~\cite{geant4} software package, which includes the geometric description of the BESIII detector~\cite{detvis, Li:2024pox, Song:2025pnt, geo1, geo2} and the detector response, are used to determine detection efficiencies and to estimate backgrounds. The simulation models the beam energy spread and initial state radiation in the $e^+e^-$ annihilations with the generator {\sc kkmc}~\cite{ref:kkmc, ref:kkmc2}. The inclusive MC sample includes both the production of the $\jpsi$ resonance and the continuum processes incorporated in {\sc kkmc}~\cite{ref:kkmc, ref:kkmc2}. All particle decays are modeled with {\sc evtgen}~\cite{ref:evtgen, ref:evtgen2} using BFs  either taken from the Particle Data Group~(PDG)~\cite{pdg:2022}, when available, or otherwise estimated with {\sc lundcharm}~\cite{ref:lundcharm, ref:lundcharm2}. Final state radiation from charged final state particles is incorporated using the {\sc photos} package~\cite{photos2}. 
The signal MC sample for the decay $\jpsi\to D_{s}^{-}\rho^{+}$ is generated with the VVS\_PWAVE model~\cite{ref:evtgen, ref:evtgen2} for the initial decay while the $\jpsi\to D_{s}^{-}\pi^{+}$ sample is generated with the VSS model~\cite{ref:evtgen, ref:evtgen2}.  The subsequent $D_{s}^{-}\to\phi e^{-}\Bar{\nu}_{e}$ decay is generated with the PHOTOS ISGW2 model while $\phi\to K^{+}K^{-}$ and $\rho^{+}\to\pi^{+}\pi^{0}$ use the VSS model~\cite{ref:evtgen, ref:evtgen2}.  

\section{EVENT SELECTION AND DATA ANALYSIS}
\label{sec:analysis}
\hspace{1.5em} 
The analysis is conducted using the BESIII offline software system~\cite{bes3:boss705}. 
Full reconstruction of $D_{s}$ mesons with non-leptonic-decay modes does not offer good sensitivity due to the large hadronic background from $\jpsi$ inclusive decays. Therefore, $D_{s}$ candidates are reconstructed via the semi-leptonic decay $D_{s}^{-}\to \phi e^{-} \Bar{\nu}_{e}$ with $\phi\to K^{+}K^{-}$, which has a large BF of $(2.39\pm0.16)\%$ and low background~\cite{pdg:2022}. For the $\rho^{+}$ side, the subsequent signal decays are $\rho^{+}\to \pi^{+}\pi^{0}$, $\pi^{0}\to\gamma\gamma$. For both decays $\jpsi\to D_{s}^{-}\rho^{+}$ and $\jpsi\to D_{s}^{-}\pi^{+}$, the net charge of the tracks $\sum Q=0$ and the number of charged tracks must be four. Four charged tracks in both decays must be detected in the MDC within the range of $|\text{cos}\theta|<0.93$, where $\theta$ is the polar angle with respect to the axis of symmetry of the MDC. The distance of closest approach from the interaction point is required to satisfy $R_{z}<$10~cm along the beam direction, and $R_{xy}<$1~cm in the plane perpendicular to the beam axis. 

Particle identification (PID) is applied by combining the measurements of d$E$/d$x$ in the MDC and the time-of-flight in the TOF to obtain the likelihoods for each hadron hypothesis $\mathcal{L}_{h}$~($h=K, \pi, e$). 
The $\pi$ candidates require $\mathcal{L}(\pi)>0$, $\mathcal{L}(\pi)>\mathcal{L}(K)$, and $\mathcal{L}(\pi)>\mathcal{L}(e)$. 
The $K$ candidates require $\mathcal{L}(K)>0$ and $\mathcal{L}(K)>\mathcal{L}(\pi)$. 
The $e$ candidates require $\mathcal{L}(e)>0.001$ and $\mathcal{L}(e)/(\mathcal{L}(e)+\mathcal{L}(\pi)+\mathcal{L}(K))>0.8$. 
One $\pi$ candidate, two $K$ candidates, and one $e$ candidate are required for each event. 

Photon candidates are selected from showers deposited in the EMC with energies of $E_{\gamma}>25\mev$ in the barrel ($|\text{cos}\theta|<0.80$) and $E_{\gamma}>50\mev$ in the end-cap ($0.86<|\text{cos}\theta|<0.92$).  Showers must be separated from the extrapolated positions of any charged track by at least $10^{\circ}$ and occur within 700~ns after the event start time.  Events with at least two photon candidates are kept in the selection of $\jpsi\to D_{s}^{-} \rho^{+}$. 

The $D_{s}^{-}\to \phi e^{-} \Bar{\nu}_{e}$ reconstruction requires one $e$ candidate and one $\phi$ candidate. To suppress the backgrounds from hadron misidentification, the following requirements for $e^{-}$ candidates are used: $0.83<E/p<1.11$ for $\jpsi\to D_{s}^{-}\rho^{+}$ and $0.87<E/p<1.14$ for $\jpsi\to D_{s}^{-}\pi^{+}$. Here $E/p$ is the ratio of the EMC deposited energy, $E$, and the MDC momentum, $p$; these criteria have been optimized according to the Punzi significance~\cite{Punzi:2003bu}. One $K^{+}$ and one $K^{-}$ candidate are used to reconstruct the $\phi$ candidate, whose invariant mass, $M_{KK}$, must lie within $(1.01, 1.03)\gevcc$. 

For $\jpsi\to D_{s}^{-} \rho^{+}$, the $\rho^{+}$ candidates are reconstructed from one $\pi^{+}$ candidate and one $\pi^{0}$ candidate. The $\pi^{0}$ candidates are selected by looping over all pairs of good photon candidates to select the one with the minimum $\chi^2_{\gamma\gamma}$ from a kinematic fit constraining their invariant mass to the known $\pi^{0}$ mass~\cite{pdg:2022}; this minimum must also satisfy $\chi^2_{\gamma\gamma} < 200$. The updated $\pi^0$ four-vector from the fit is used in later kinematic calculations. To select the $\rho^+$ candidates, $|M_{\pi\pi}-M_{\rho}^{\rm{PDG}}|<0.15\gevcc$ is required, where $M_{\pi\pi}$ is the $\pi^{+}\pi^{0}$ invariant mass and $M_{\rho}^{\rm{PDG}}$ is the PDG resonance mass~\cite{pdg:2022}. The total energy~($E_{\gamma}^{\rm{rest}}$) of photons, except for the two photons that are used to reconstruct the $\pi^{0}$, must be less than $0.22\gev$ to reduce backgrounds with extra photons.

Not directly detectable, neutrinos may be inferred from the missing energy and missing momentum. To suppress the backgrounds from $\jpsi$ hadronic decays without a missing particle, we require the missing momentum $|\vec{p}_{\rm{miss}}|=|\vec{p}_{\jpsi}-\vec{p}_{K^{+}}-\vec{p}_{K^{-}}-\vec{p}_{e^{-}}-\vec{p}_{\rho^+(\pi^+)}|$ to be larger than 0.15$\gevc$ for $\jpsi\to D_{s}^{-}\rho^{+}$ and 0.10$\gevc$ for $\jpsi\to D_{s}^{-}\pi^{+}$, where $\vec{p}_{\jpsi}$, $\vec{p}_{K^{+}}$, $\vec{p}_{K^{-}}$, $\vec{p}_{e^{-}}$ and $\vec{p}_{\rho^{+}(\pi^{+})}$ are the momenta of $\jpsi$, $K^{+}$, $K^{-}$, $e^{-}$, and $\rho^+$~($\pi^{+}$), respectively. To remove the background from $\jpsi\to K^{+}K^{-}\pi^{+}\pi^{-}$ and multi-$\pi^{0}/\gamma$ in final states, we apply a requirement $|\text{U}_{\rm{miss}}| < 0.04 \gev$, with $\text{U}_{\rm{miss}}$ defined as:
\begin{eqnarray}
\text{U}_{\rm{miss}}=E_{\rm{miss}}-|\vec{p}_{\rm{miss}}|c, 
\end{eqnarray}
\begin{eqnarray}
E_{\rm{miss}} = E_{\jpsi}-E_{K^{+}}-E_{K^{-}}-E_{e^{-}}-E_{\rho^{+}(\pi^{+})},
\end{eqnarray}
 where $E_{\jpsi}$, $E_{K^{+}}$, $E_{K^{-}}$, $E_{e^{-}}$, $E_{K^{+}}$, and $E_{\rho^{+}(\pi^{+})}$ are the energies of $\jpsi$, $K^{+}$, $K^{-}$, $e^{-}$, and $\rho^+$~($\pi^{+}$), respectively.

After the above selection, most background events are removed according to inclusive MC sample results. To study the remaining background sources, we analyze the MC truth information for background events passing the selections. Most background events contain a $\pi^-$ being misidentified as an $e^-$ candidate. Thus, we further require for $e^{-}$ candidates that $\chi_{\pi}^{\text{track}~e}<3.0$ and $\chi_{e}^{\text{track}~e}>-1.0$ for $\jpsi\to D_{s}^{-}\rho^{+}$, and $\chi_{\pi}^{\text{track}~e}<2.6$ for $\jpsi\to D_{s}^{-}\pi^{+}$, in which $\chi_{\pi(e)}^{\text{track}~e}$ is the normalized deviation of the measured d$E$/d$x$ from the expected value under $\pi$~($e$) hypothesis. These requirements are optimized according to the Punzi significance~\cite{Punzi:2003bu}.


The $D_{s}^{-}$ candidates are identified by the mass recoiling against the $\rho^{+}$~($\pi^{+}$) candidates.  Specifically, the signal variable is 
$M_{D_{s}}$ as the signature of signal events, which is defined by 
 \begin{eqnarray}
 M_{D_{s}}=\sqrt{E_{D_{s}}^2/c^{4}-|\Vec{p}_{D_{s}}|^{2}/c^{2}},
 \end{eqnarray}
{\rm with} 
  \begin{eqnarray}
 E_{D_{s}}=E_{\jpsi}-E_{\rho^{+}(\pi^{+})} \, , \quad
 \Vec{p}_{D_{s}}=\Vec{p}_{\jpsi}-\Vec{p}_{\rho^{+}(\pi^{+})}.
 \end{eqnarray}
 The signal will appear as a peak with a central value around the known $D_{s}$ mass~\cite{pdg:2022}.  

\vspace{-0.0cm}
\begin{figure*}[htbp] \centering
	\setlength{\abovecaptionskip}{-1pt}
	\setlength{\belowcaptionskip}{10pt}
 
        \subfigure[]
        {\includegraphics[width=0.49\textwidth]{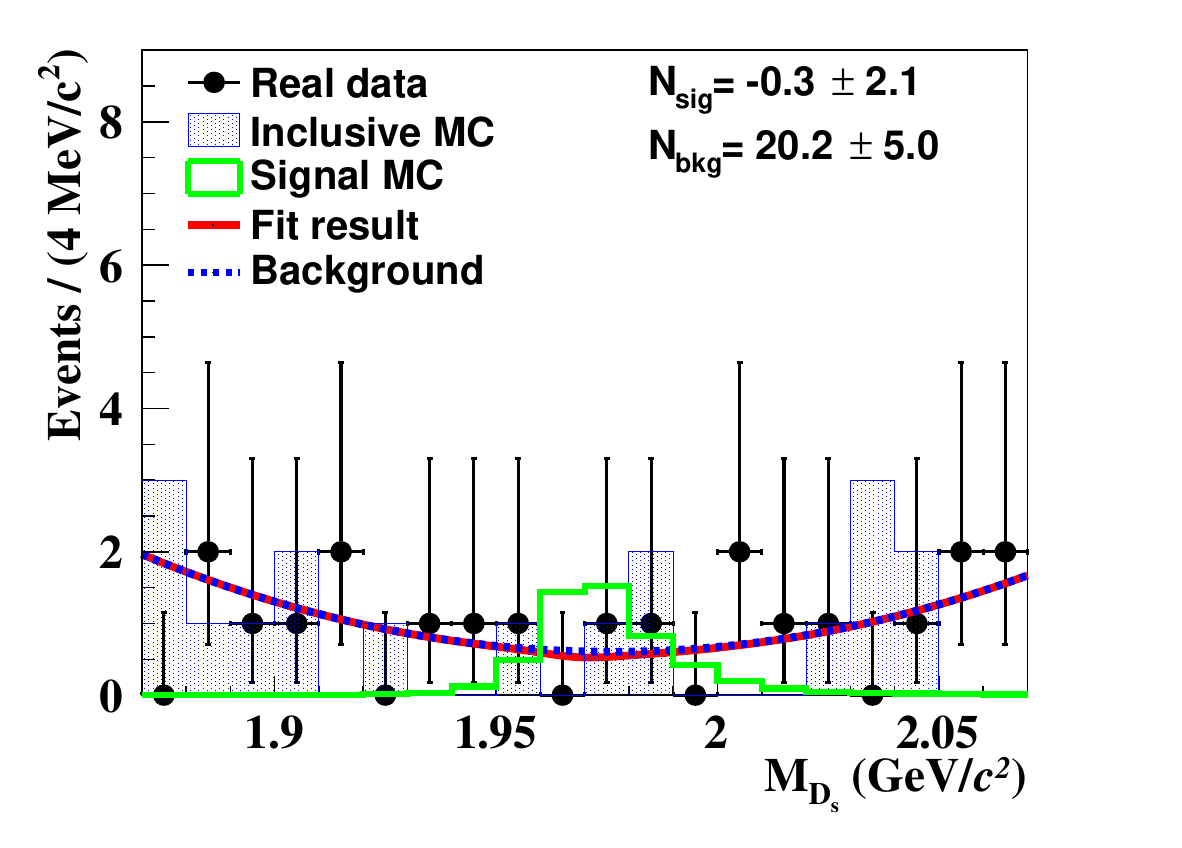}}
        \subfigure[]
        {\includegraphics[width=0.49\textwidth]{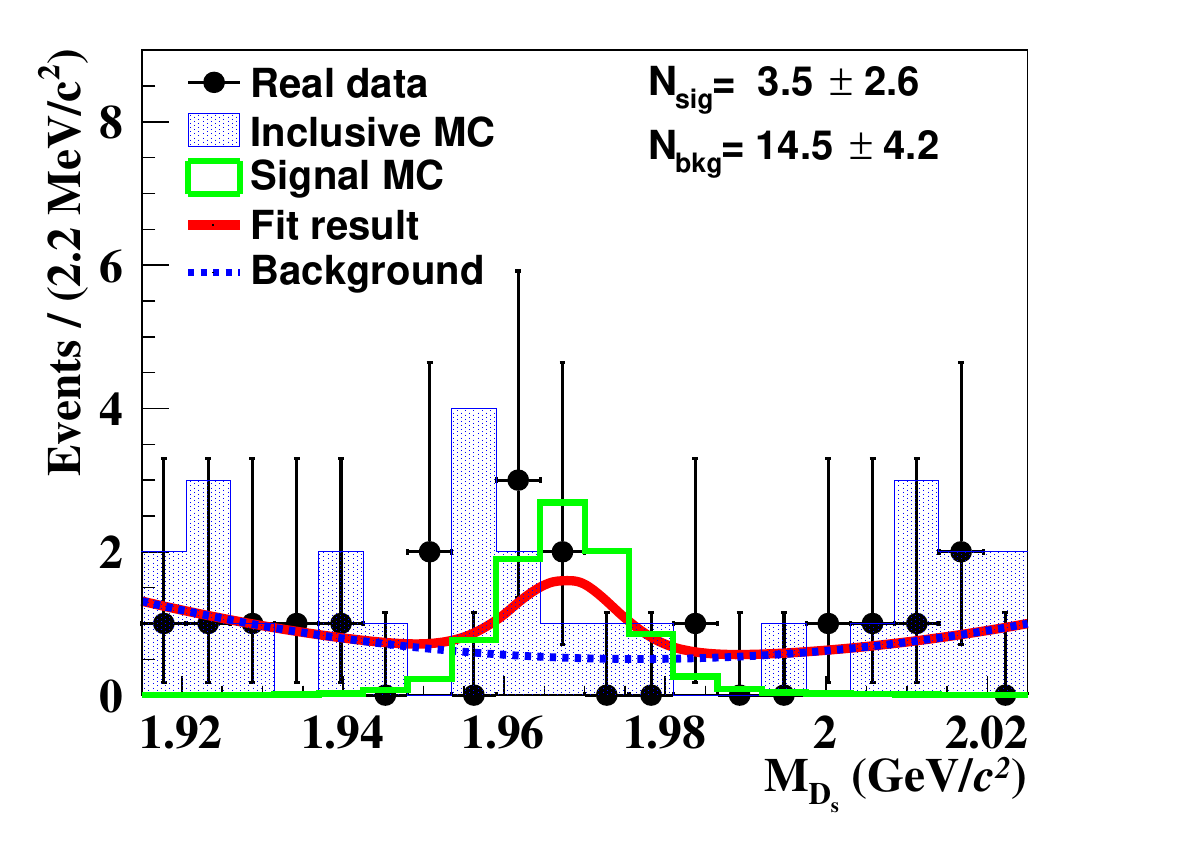}}\\
        
	\caption{The $M_{D_{s}}$ distributions of the (a) $\jpsi\to D_{s}^{-}\rho^{+}$ and (b) $\jpsi\to D_{s}^{-}\pi^{+}$ candidate events. The black dots with error bars are data; the blue histograms are the inclusive MC samples; the green solid lines are the signal MC samples, scaled to the obtained 90\% UL BFs. The red line is the total fit result with the signal shape and the blue background shape dotted line.} 
	\label{fig:2fit}
\end{figure*}
\vspace{-0.0cm}

\section{RESULT}
\label{sec:result}
\hspace{1.5em}
Unbinned extended maximum likelihood fits are performed to the distributions of $M_{D_{s}}$ to determine the signal yields, as shown in Fig.~\ref{fig:2fit}. 
The signal probability density function (PDF) is derived from the shape of signal MC simulation. The background shape is described with a second-order polynomial function.

The results indicate no significant excess of observed signal above the backgrounds.
Therefore, we set the ULs on $\mathcal{B}(\jpsi\to D_{s}^{-}\rho^{+})$ and $\mathcal{B}(\jpsi\to D_{s}^{-}\pi^{+})$ at the $90\%$ C.L. after considering the systematic uncertainties.
The branching fractions $\mathcal{B}$ is calculated by 
\begin{eqnarray}
\mathcal{B}=\frac{\text{N}_{\rm{sig}}}{\text{N}_{\jpsi} \, \epsilon \, \mathcal{B}_{\rm{inter}}},
\label{eq:pdf}
\end{eqnarray}
where $\text{N}_{\rm{sig}}$, $\text{N}_{\jpsi}$, $\epsilon$, and $\mathcal{B}_{\rm{inter}}$ are the fitted signal yield, the total number of $\jpsi$ events, the signal efficiency, and the intermediate decay BF, respectively.
We scan the $\jpsi\to D_{s}^{-}\rho^{+}$~($\jpsi\to D_{s}^{-}\pi^{+}$) signal yields 200~(300) times by varying the number of events with a step size of 0.1 to obtain the likelihood $\mathcal{L}_{i}$ at each step ${i}$. The relative likelihood $\mathcal{L}$ is defined as
\begin{eqnarray}
\mathcal{L}=\frac{\mathcal{L}_{i}}{\mathcal{L}_{\rm{max}}},
\label{eq:L}
\end{eqnarray}
where $\mathcal{L}_{\rm{max}}$ is the maximum value of all $\mathcal{L}_{i}$ in the scan. A Gaussian fit is performed to the dependence of the relative likelihood on the BF. The fitting function is 
\begin{eqnarray}
\mathcal{L}(\mathcal{B})_{\rm{fit}}\propto \rm{exp} \left[-\frac{(\mathcal{B}-\hat{\mathcal{B}})^{2}}{2\sigma^{2}_{\mathcal{B}}}\right], 
\label{eq:B}
\end{eqnarray}
where $\hat{\mathcal{B}}$ and $\sigma_{\mathcal{B}}$ are the fitted mean value and uncertainty of the BF. 
Following a method which incorporates the systematic uncertainties into the UL of the BF~\cite{Liu:2015uha}, we obtain the smeared likelihood function by convolving with a Gaussian function:
\begin{eqnarray}
\mathcal{L}(\mathcal{B})_{\rm{smear}}\propto \int_{0}^{1}\rm{exp}\left[-\frac{(\epsilon\mathcal{B}/\hat{\epsilon}-\hat{\mathcal{B}})^{2}}{2\sigma^{2}_{\mathcal{B}}}\right] \, \frac{1}{\sqrt{2\pi}\sigma_{\epsilon}} \, \rm{exp}\left[-\frac{(\epsilon-\hat{\epsilon})^{2}}{2\sigma^{2}_{\epsilon}}\right] \, d\epsilon,
\label{eq:B}
\end{eqnarray}
where $\hat{\epsilon}$ is the nominal efficiency, $\sigma_{\epsilon}=\sigma_{\rm{sys}} \, \epsilon$ is the systematic uncertainty of efficiency. For the $D_s^-\pi^+$ channel, the $\mathcal{L}(\mathcal{B})_{\rm{fit}}$ distribution is asymmetrical and is fitted with a double Gaussian before smearing is applied.  

\begin{figure*}[htbp] \centering
	\setlength{\abovecaptionskip}{-1pt}
	\setlength{\belowcaptionskip}{10pt}
 
        \subfigure[]
        {\includegraphics[width=0.49\textwidth]{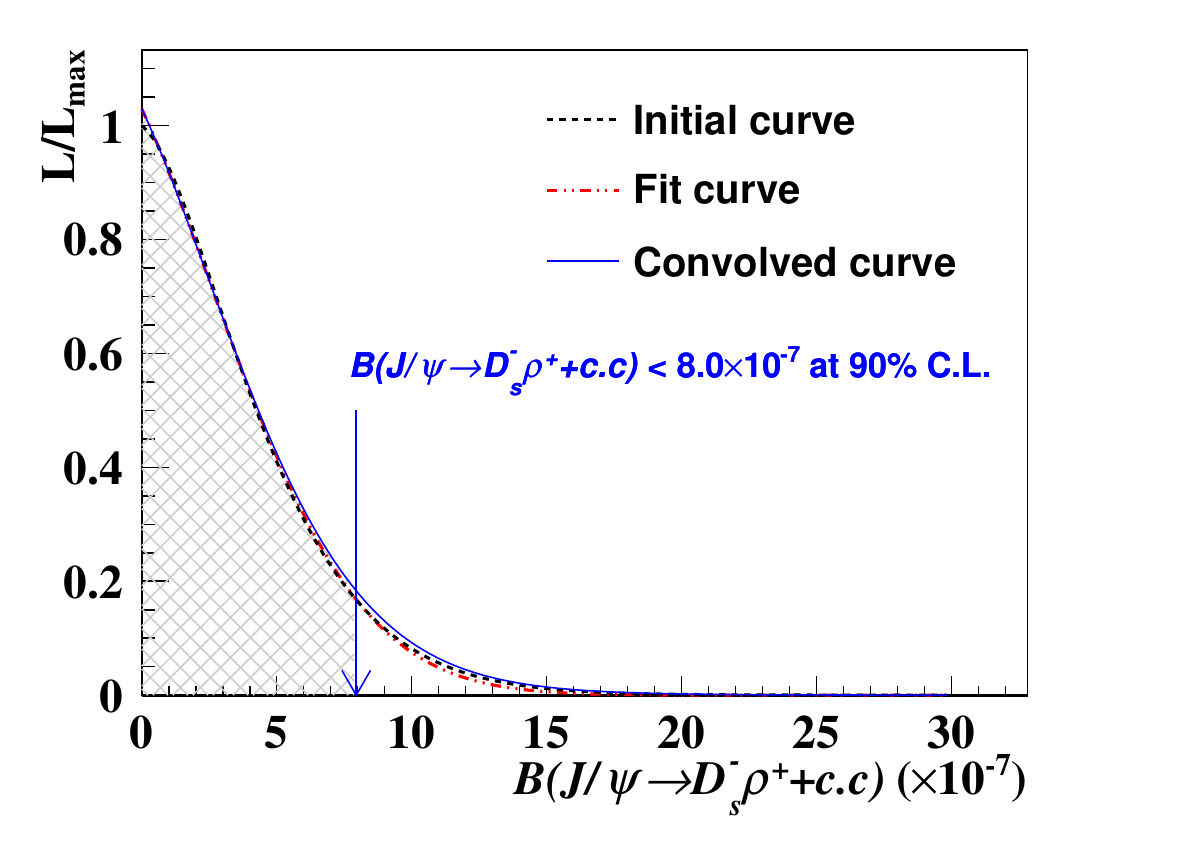}}
        \subfigure[]
        {\includegraphics[width=0.49\textwidth]{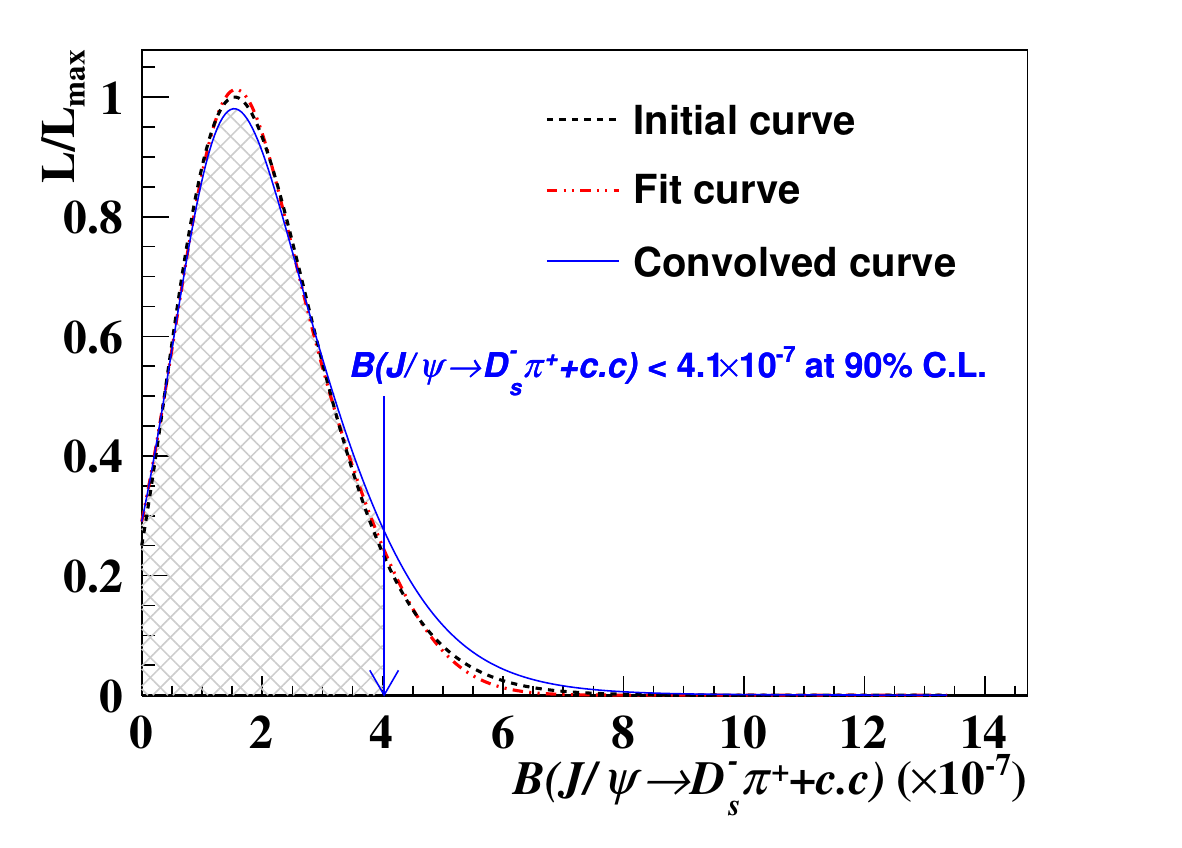}}\\
        
	\caption{The distributions of the likelihood scan values for (a) $\jpsi\to D_{s}^{-}\rho^{+}$ and (b) $\jpsi\to D_{s}^{-}\pi^{+}$. The dotted-dashed black line is the initial curve, the dotted-dashed red line is the fit curve with a Gaussian function, and the solid blue line is the fit curve convolved with a Gaussian for systematic uncertainties. The blue arrows indicate the ULs on BFs at the 90\% C.L.} 
\label{fig:limit}
\end{figure*}

To ensure a conservative assessment of the impact of fluctuations resulting from the fit range of $M_{D_{s}}$, we have systematically varied the fit ranges multiple times. Additionally, to account for the influence of fluctuations introduced by the background fit shape of $M_{D_{s}}$, we have employed a first-order polynomial function to replace the initial second-order polynomial function. We choose the most conservative ULs $\mathcal{B}(\jpsi\to D_{s}^{-}\rho^{+}+c.c.)<8.0\times10^{-7}$ and $\mathcal{B}(\jpsi\to D_{s}^{-}\pi^{+}+c.c.)<4.1\times10^{-7}$ at the $90\%$ C.L., as shown in Fig.~\ref{fig:limit}.
\vspace{-0.0cm}
\vspace{-0.0cm}

\section{SYSTEMATIC UNCERTAINTY}
\label{sec:systematic}
\hspace{1.5em}
The systematic uncertainties in the BF measurement of $\jpsi\to D_{s}^{-}\rho^{+}$ and $\jpsi\to D_{s}^{-}\pi^{+}$ mainly come from the signal MC model, particle tracking and PID, intermediate BFs, the total number of $\jpsi$ events, MC statistics, and the requirements imposed on $M_{\phi}$,   $|\vec{p}_{\rm{miss}}|$, $\text{U}_{\rm{miss}}$, $E/p$, and $\chi_{\pi}^{\text{track}~e}$. In addition, photon detection efficiency and the requirements on $\chi_{e}^{\text{track}~e}$, $M_{\rho}$, and $E^{\rm{rest}}_{\gamma}$ are sources of the systematic uncertainties in the BF measurement of $\jpsi\to D_{s}^{-}\rho^{+}$. The influence of fluctuations introduced by the background fit shape of $M_{D_{s}}$ is discussed in Section~\ref{sec:result}. 
The systematic uncertainties from different sources are studied in the following items and summarized in Table~\ref{tab:syst_err}. 
The quadratic sum of all the uncertainties is taken as the total systematic uncertainty. 

\begin{itemize}
    \item \emph{MC generator model.} To estimate the systematic uncertainty due to the MC generator model, signal MC samples generated with alternative phase space (PHSP) models are compared to the nominal signal MC samples. The efficiency differences between the alternative and nominal signal MC samples are assigned as the systematic uncertainties: 10.5\% and 13.1\%, for $\jpsi\to D_{s}^{-}\rho^{+}$ and $\jpsi\to D_{s}^{-}\pi^{+}$, respectively.

    \item \emph{Tracking and PID efficiencies.} 
    The uncertainties of the tracking and PID efficiencies for $K^{\pm}$ and $\pi^{\pm}$ have been studied by analyzing $\psi(3770)\to D^{0}\Bar{D}^{0}$ ($D^{+}D^{-}$) decay events. The hadronic decays of $D^{0}\to K^{-}\pi^{+}$, $K^{-}\pi^{+}\pi^{+}\pi^{-}$ versus $\Bar{D}^{0}\to K^{+}\pi^{-}$, $K^{+}\pi^{-}\pi^{-}\pi^{+}$, and $D^{+}\to K^{-}\pi^{+}\pi^{+}$ versus $D^{-}\to K^{+}\pi^{-}\pi^{-}$, are used as control samples for missing-particle studies.  The electron tracking and PID efficiencies of the datasets are studied using control samples of the process $ e^{+} e^{-} \to e^{+} e^{-} \gamma$~(including $\jpsi\to e^{+}e^{-}\gamma$) from MC simulation at $3.097\gev$ and the corresponding data. We calculate the systematic uncertainties of $K, \pi, e$ tracking and $K, \pi, e$ PID, accounting for the momentum distributions of $K, \pi, e$ in data.   
 The total systematic uncertainties from tracking are 2.6\% for $\jpsi\to D_{s}^{-}\rho^{+}$ and 2.7\% for $\jpsi\to D_{s}^{-}\pi^{+}$. The total systematic uncertainties from PID are 1.1\% for $\jpsi\to D_{s}^{-}\rho^{+}$ and 1.3\% for $\jpsi\to D_{s}^{-}\pi^{+}$.

     \item \emph{Intermediate BFs.} The uncertainty of the  intermediate decay BFs quoted from the PDG~\cite{pdg:2022} is 6.8\% for both $\jpsi\to D_{s}^{-}\rho^{+}$ and $\jpsi\to D_{s}^{-}\pi^{+}$.

     \item \emph{Number of $\jpsi$ events.} The total number of $\jpsi$ events is $\text{N}_{\jpsi}=(10087\pm44)\times10^{6}$~\cite{bes3:totJpsiNumber}, where the uncertainty is systematic and the statistical one is negligible. We take $0.5\%$ as the systematic uncertainty.

     \item \emph{MC statistics.}  The systematic uncertainties due to MC statistics are 0.6\% for $\jpsi\to D_{s}^{-}\rho^{+}$ and 0.4\% for $\jpsi\to D_{s}^{-}\pi^{+}$.

    \item \emph{$M_{\phi}$ requirement.} The systematic uncertainty of the $M_{\phi}$ requirement is estimated as 5.2\% via a control sample of $\jpsi\to\phi \phi\gamma$ with $\phi\to K^{+}K^{-}$. 

    \item \emph{$|\vec{p}_{\rm{miss}}|$, $\text{U}_{\rm{miss}}$, $E/p$,  $\chi_{\pi}^{\text{track}~e}$, and $\chi_{e}^{\text{track}~e}$ requirements.} To estimate the systematic uncertainties caused by the requirements on these quantities, we use a control sample of $D^{0}\to K^{+}e^{-}\Bar{\nu}$. The systematic uncertainties for $\jpsi\to D_{s}^{-}\rho^{+}$ are $4.1\%$, $5.7\%$, $1.5\%$, $5.7\%$, and $2.9\%$, respectively. For $\jpsi\to D_{s}^{-}\pi^{+}$ only the first four quantities are relevant and the obtained uncertainties are $2.6\%$, $5.7\%$, $1.8\%$, and $5.0\%$.

    \item \emph{$M_{\rho}$ and $E_{\gamma}^{\rm{rest}}$ requirements.} The systematic uncertainties on the $M_{\rho}$ and $E_{\gamma}^{\rm{rest}}$ requirements are estimated as 2.8\% and 3.3\% using a control sample of $\jpsi\to\rho^{+}\pi^{-}$ with $\rho^{+}\to\pi^{+}\pi^{0}$. 

    \item \emph{$\gamma$ detection.} The uncertainty due to $\gamma$ detection is assigned as $1\%$ per $\gamma$ by using $\jpsi\to \rho^{0}\pi^{0}$ and $e^{+}e^{-}\to\gamma\gamma$ samples~\cite{BESIII:2010ank}.

\end{itemize}
The total systematic uncertainties are $17.5\%$ for $\jpsi\to D_{s}^{-}\rho^{+}$ and $17.9\%$ for $\jpsi\to D_{s}^{-}\pi^{+}$.
\begin{table*}[tpb]
\setlength{\abovecaptionskip}{0.0cm}
\setlength{\belowcaptionskip}{-1.6cm}
\caption{Summary of the systematic uncertainties for the BF measurements. 
A ``$-$'' indicates a non-applicable source.  
The total value is calculated by summing up all sources in quadrature.}
  \begin{center}
  \footnotesize
  \newcommand{\tabincell}[2]{\begin{tabular}{@{}#1@{}}#2\end{tabular}}
  \begin{threeparttable}
  \begin{tabular}{l|c c c c c c c}
      \hline\hline
                         Sources &  $\Delta_{sys}$($\jpsi\to D_{s}\rho$) ($\%$) & $\Delta_{sys}$($\jpsi\to D_{s}\pi$) ($\%$)\\
	\hline
	MC generator model        & 10.5 & 13.1 \\
	Tracking                  &  2.6 &  2.7 \\
	Particle ID               &  1.1 &  1.3 \\
	Intermediate BFs          &  6.8 &  6.8 \\
	Total number of $J/\psi$  &  0.5 &  0.5 \\
        MC statistics             &  0.6 &  0.4 \\
	$M_{\phi}$ requirement    &  5.2 &  5.2 \\
	$|\vec{p}_{\rm{miss}}|$ requirement     & 4.1 & 2.6 \\
	$\text{U}_{\rm{miss}}$ requirement      & 5.7 & 5.7 \\
	$E/p$ requirement         &  1.5 &  1.8 \\
	$\chi_{\pi}^{\rm{track}~e}$ requirement & 5.7 & 5.0 \\
        $\chi_{e}^{\rm{track}~e}$ requirement   & 2.9 & $-$ \\
	$M_{\rho}$ requirement    &  2.8 &  $-$ \\
        $E_{\gamma}^{\rm{rest}}$ requirement    & 3.3 & $-$ \\
        $\gamma$ detection        &  2.0 &  $-$ \\
        \hline
        Total                     & 17.5 & 17.9 \\
      \hline\hline
  \end{tabular}
  \label{tab:syst_err}
  \end{threeparttable}
  \end{center}
\end{table*}
\newpage
\section{SUMMARY}
\label{sec:summary}
\hspace{1.5em}
The rare charmonium weak decays $\jpsi\to D_{s}^{-}\rho^{+}$ and $\jpsi\to D_{s}^{-}\pi^{+}$ are searched for based on $(10087\pm44)\times10^{6}$ $\jpsi$ events collected with the BESIII detector.  
No significant signal is observed and ULs on their BFs are set at $\mathcal{B}(\jpsi\to D_{s}^{-}\rho^{+})<8.0\times10^{-7}$ and $\mathcal{B}(\jpsi\to D_{s}^{-}\pi^{+})<4.1\times10^{-7}$ at the 90\% C.L. In comparison to the previous best limits, the UL for $\jpsi\to D_{s}^{-}\rho^{+}$ has been improved by about an order of magnitude, and the UL for $\jpsi\to D_{s}^{-}\pi^{+}$ has been improved by about three orders of magnitude. These results are consistent with the SM-based predictions~\cite{wang:2008a, shen:2008, dhir:2013, Sun:2023uyn}.

\acknowledgments
\hspace{1.5em}
The BESIII Collaboration thanks the staff of BEPCII (https://cstr.cn/31109.02.BEPC) and the IHEP computing center for their strong support. This work is supported in part by National Key R\&D Program of China under Contracts Nos. 2023YFA1606000, 2023YFA1606704; National Natural Science Foundation of China (NSFC) under Contracts Nos. 11635010, 11935015, 11935016, 11935018, 12025502, 12035009, 12035013, 12061131003, 12192260, 12192261, 12192262, 12192263, 12192264, 12192265, 12221005, 12225509, 12235017, 12361141819; the Chinese Academy of Sciences (CAS) Large-Scale Scientific Facility Program; the Strategic Priority Research Program of Chinese Academy of Sciences under Contract No. XDA0480600; CAS under Contract No. YSBR-101; 100 Talents Program of CAS; The Institute of Nuclear and Particle Physics (INPAC) and Shanghai Key Laboratory for Particle Physics and Cosmology; Agencia Nacional de Investigación y Desarrollo de Chile (ANID), Chile under Contract No. ANID PIA/APOYO AFB230003; ERC under Contract No. 758462; German Research Foundation DFG under Contract No. FOR5327; Istituto Nazionale di Fisica Nucleare, Italy; Knut and Alice Wallenberg Foundation under Contracts Nos. 2021.0174, 2021.0299; Ministry of Development of Turkey under Contract No. DPT2006K-120470; National Research Foundation of Korea under Contract No. NRF-2022R1A2C1092335; National Science and Technology fund of Mongolia; Polish National Science Centre under Contract No. 2024/53/B/ST2/00975; STFC (United Kingdom); Swedish Research Council under Contract No. 2019.04595; U. S. Department of Energy under Contract No. DE-FG02-05ER41374

\newpage
M.~Ablikim$^{1}$\BESIIIorcid{0000-0002-3935-619X},
M.~N.~Achasov$^{4,b}$\BESIIIorcid{0000-0002-9400-8622},
P.~Adlarson$^{77}$\BESIIIorcid{0000-0001-6280-3851},
X.~C.~Ai$^{82}$\BESIIIorcid{0000-0003-3856-2415},
R.~Aliberti$^{36}$\BESIIIorcid{0000-0003-3500-4012},
A.~Amoroso$^{76A,76C}$\BESIIIorcid{0000-0002-3095-8610},
Q.~An$^{73,59,\dagger}$,
Y.~Bai$^{58}$\BESIIIorcid{0000-0001-6593-5665},
O.~Bakina$^{37}$\BESIIIorcid{0009-0005-0719-7461},
Y.~Ban$^{47,g}$\BESIIIorcid{0000-0002-1912-0374}
% ...
\\
\vspace{0.2cm}
(BESIII Collaboration)\\
\vspace{0.2cm} {\it
$^{1}$ Institute of High Energy Physics, Beijing 100049, People's Republic of China\\
% ...
$^{4}$ Budker Institute of Nuclear Physics SB RAS (BINP), Novosibirsk 630090, Russia\\
% ...
$^{36}$ Johannes Gutenberg University of Mainz, Johann-Joachim-Becher-Weg 45, D-55099 Mainz, Germany\\
$^{37}$ Joint Institute for Nuclear Research, 141980 Dubna, Moscow region, Russia\\
% ...
$^{47}$ Peking University, Beijing 100871, People's Republic of China\\
% ...
$^{58}$ Southeast University, Nanjing 211100, People's Republic of China\\
$^{59}$ State Key Laboratory of Particle Detection and Electronics, Beijing 100049, Hefei 230026, People's Republic of China\\
% ...
$^{73}$ University of Science and Technology of China, Hefei 230026, People's Republic of China\\
% ...
$^{76}$ University of Turin and INFN, (A)University of Turin, I-10125, Turin, Italy; (B)University of Eastern Piedmont, I-15121, Alessandria, Italy; (C)INFN, I-10125, Turin, Italy\\
$^{77}$ Uppsala University, Box 516, SE-75120 Uppsala, Sweden\\
% ...
$^{82}$ Zhengzhou University, Zhengzhou 450001, People's Republic of China\\

\vspace{0.2cm}
$^{\dagger}$ Deceased\\
% ...
$^{b}$ Also at the Novosibirsk State University, Novosibirsk, 630090, Russia\\
% ...
$^{g}$ Also at State Key Laboratory of Nuclear Physics and Technology, Peking University, Beijing 100871, People's Republic of China\\
% ...

}

\end{document}